\documentclass[aps, prb, twocolumn, superscriptaddress, floatfix, longbibliography, showkeys, 10pt]{revtex4-2}

\usepackage{graphicx}
\usepackage{amsmath, amssymb}
\usepackage{braket}
\usepackage{chemformula}
\usepackage[separate-uncertainty=true]{siunitx}
\usepackage{tikz}
\usetikzlibrary{matrix,positioning,arrows,arrows.meta}
\usepackage{tikz-3dplot}
\usepackage{mleftright}

\makeatletter
\def\@hangfrom@section#1#2#3{\@hangfrom{#1#2}#3}
\def\@hangfroms@section#1#2{#1#2}
\def\@seccntformat#1{\csname the#1\endcsname\quad}
\def\section{\@startsection
  {section}{1}{\z@}%
  {-2.8ex \@plus -1ex \@minus -.2ex}%
  {1.5ex \@plus .2ex}%
  {\normalfont\large\bfseries\raggedright}}

\def\subsection{\@startsection
  {subsection}{2}{\z@}%
  {-2.2ex \@plus -1ex \@minus -.2ex}%
  {1ex \@plus .2ex}%
  {\normalfont\normalsize\bfseries\raggedright}}

\makeatother
\newcommand{\fcite}[1]{\textsuperscript{\cite{#1}}}
\makeatletter



\usepackage[colorlinks=true,
            linkcolor=blue,
            citecolor=blue,
            urlcolor=blue]{hyperref}
\usepackage{cleveref}
\usepackage{tabularray}
\UseTblrLibrary{booktabs}
\usepackage[T1]{fontenc}
\usepackage{float}
\usepackage{blkarray}
\usepackage{enumitem}
\usepackage[utf8]{inputenc}

\newcommand*{\mixperov}{\ch{FA_{0.9}Cs_{0.1}PbI_{2.8}Br_{0.2}}}
\newcommand*{\BDSS}{\Delta_{\text{X}}}

\begin{document}

\title{Bright-dark exciton splitting in lead halide perovskite crystals\\ accessed via quantum beats in photon echoes}
\author{M.~Alex~Hollberg}
\affiliation{Experimentelle Physik 2, Technische Universit\"at Dortmund, 44227 Dortmund, Germany}
\author{Mikhail.~O.~Nestoklon}
\affiliation{Experimentelle Physik 2, Technische Universit\"at Dortmund, 44227 Dortmund, Germany}
\author{Artur~V.~Trifonov}
\affiliation{Experimentelle Physik 2, Technische Universit\"at Dortmund, 44227 Dortmund, Germany}
\author{Stefan~Grisard}
\affiliation{Experimentelle Physik 2, Technische Universit\"at Dortmund, 44227 Dortmund, Germany}
\author{Oleh Hordiichuk}
\affiliation{Laboratory of Inorganic Chemistry, Department of Chemistry and Applied Biosciences, ETH Zürich, CH-8093 Zürich, Switzerland}
\affiliation{Laboratory for Thin Films and Photovoltaics, Empa-Swiss Federal Laboratories for Materials Science and Technology, CH-8600 Dübendorf, Switzerland}
\author{Dmitry N. Dirin}
\affiliation{Laboratory of Inorganic Chemistry, Department of Chemistry and Applied Biosciences, ETH Zürich, CH-8093 Zürich, Switzerland}
\affiliation{Laboratory for Thin Films and Photovoltaics, Empa-Swiss Federal Laboratories for Materials Science and Technology, CH-8600 Dübendorf, Switzerland}
\author{Maksym V. Kovalenko}
\affiliation{Laboratory of Inorganic Chemistry, Department of Chemistry and Applied Biosciences, ETH Zürich, CH-8093 Zürich, Switzerland}
\affiliation{Laboratory for Thin Films and Photovoltaics, Empa-Swiss Federal Laboratories for Materials Science and Technology, CH-8600 Dübendorf, Switzerland}
\author{Dimitri~R.~Yakovlev}
\affiliation{Experimentelle Physik 2, Technische Universit\"at Dortmund, 44227 Dortmund, Germany}
\author{Manfred~Bayer}
\affiliation{Experimentelle Physik 2, Technische Universit\"at Dortmund, 44227 Dortmund, Germany}
\affiliation{Research Center FEMS, Technische Universit\"at Dortmund, 44227 Dortmund, Germany}
\author{Ilya~A.~Akimov}
\affiliation{Experimentelle Physik 2, Technische Universit\"at Dortmund, 44227 Dortmund, Germany}

\date{\today}

\begin{abstract}
Understanding the fine structure of excitons is crucial for optoelectronic and quantum photonic applications of lead halide perovskites. It is demonstrated that polarization-sensitive photon echo spectroscopy in magnetic field provides a powerful method to access coherent exciton dynamics and reveal their energy level structure, which is hidden by inhomogeneous broadening. Exciton quantum beats observed in both Faraday and Voigt geometries offer a precise probe of the energy splittings among the four 1$s$ exciton states, enabling determination of the fine structure and bright-dark splittings. 
Application of this technique to bulk mixed halide perovskite crystals \mixperov{} reveals a bright-dark exciton splitting of $\Delta_{\rm X} = 0.46$~meV, along with electron and hole Land\'e $g$ factors $g_\mathrm{e} = 3.38$ and $g_\mathrm{h} = -1.14$, respectively. The quantum beats persist on timescales of 20--50 ps, demonstrating remarkably robust spin and optical coherences at cryogenic temperature of 2~K. The decay of the quantum beats of the outer doublet is governed by dephasing due to dispersion of the bright-dark splitting of $\sim0.06$~meV caused by localization potential fluctuations, while dephasing in the bright exciton inner doublet originates from a small zero field splitting of $\sim0.035$~meV due to anisotropic potentials.
\end{abstract}


\maketitle

\section{Introduction}

Lead halide perovskites represent an appealing material system with a wide range of optoelectronic and photonic applications, including solar cells\fcite{24ReviewSPP,25HanNat}, photodetectors\fcite{23SakhatskyiNP,24LiuAPR}, light-emitting diodes\fcite{25RogachAM}, and lasers\fcite{23PietkaNC, 25EsmannAOM,25QihuaAM}. The energy structure of elementary optical excitations in perovskite semiconductors --- excitons --- plays a crucial role in all these applications. In particular, the fine structure splitting resulting from the exchange interaction between electrons and holes with different spin configurations governs the recombination dynamics at both cryogenic and room temperatures\fcite{18NestoklonPRB, 19TamaratNatMat, 22HanNatMat, 21CundiffSA}. Furthermore, spin-related phenomena and magnetic control of energy splitting via the Zeeman effect are of great interest for spintronic applications, as evidenced by recent advancements in electroluminescence\fcite{19VardenyNC}, giant optical orientation of excitons\fcite{24KoptevaAdvSci}, and long-lived spin polarization\fcite{19BelykhNC, 23WuNatNano}.

While the exciton spin structure has been extensively investigated in systems with narrow exciton lines, such as single lead halide perovskite nanocrystals\fcite{19TamaratNatMat, 14FerneeChSocR, 22RainoNatMat, 22HanNatMat}, it remains largerly unexplored in bulk materials. In single bulk crystals, disorder typically masks the fine structure splitting, which is hidden by strong inhomogeneous broadening of exciton emission band. Polarization-resolved measurements enable access to the fine structure of bright excitons, as demonstrated in the orthorhombic phase of MAPbBr$_3$\fcite{19BaranowskiNanoLet}. In contrast, the dark state in bulk perovskites has not yet been resolved in photoluminescence spectra. Only a few attempts have been made to estimate the energy splitting between exciton states with total angular momentum $J=0$ (dark singlet) and $J=1$ (bright triplet) in perovskite semiconductors. The bright-dark splitting $\Delta_{\rm X}$ decreases with reduced exciton confinement, i.e. with increasing nanocrystal size. In this context, extrapolation of the splitting energy for nanocrystals to infinite size, yields an estimated value of $\Delta_{\rm X}$ on the order of 0.7~meV for CsPbI$_3$\fcite{23TamaratNatCom}. Recently, Kopteva et al. observed quantum beats in the linear polarization of exciton emission in a bulk \mixperov{} crystal evidencing presence of a finite bright-dark splitting of $\Delta_{\rm X} > 0.4$~meV~\fcite{24KoptevaAdvSci}. However, to the best of our knowledge no direct measurements of this splitting have been reported in bulk crystals to date.

In this study, we address the bright-dark splitting and magnetic-field-induced Zeeman splitting of exciton states in \mixperov{} single crystals. Importantly, the mixed composition of our crystal leads to weak exciton localization, resulting in long exciton coherence time of about $T_2 \approx 57$~ps at the temperature of $T=2$~K. This corresponds to homogeneous linewidth of 23~$\mu$eV, allowing us to detect sub-meV energy splittings. At the same time, photon echo (PE) spectroscopy~\cite{22KirsteinAdvMat} effectively overcomes the inhomogeneous broadening, which is three orders of magnitude larger. 
By resolving quantum beats in the photon echo signal measured in Voigt and Faraday magnetic field configurations, we evaluate the key parameters describing the four-level structure of the 1$s$ exciton. We obtain a bright-dark splitting of $\Delta_{\rm X} = 0.46$~meV, as well as Land\'e factors of electrons $g_{\rm e} = 3.38$ and holes $g_{\rm h} = -1.14$.

\section{Results}
\label{sec:theory}

\subsection{Exciton energy structure in magnetic field}

\begin{figure*}
    \centering
    \includegraphics[width=\linewidth]{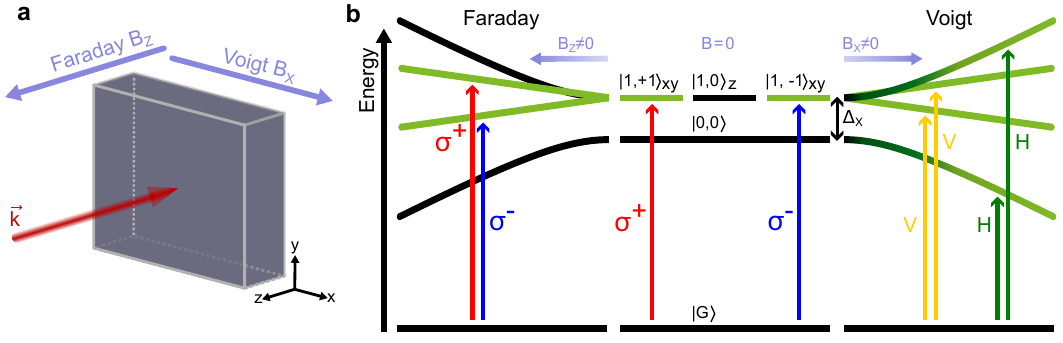}
    \caption{a) Orientation of the excitation and magnetic field geometry relative to the sample surface. The laser propagates along the $z$-axis ($\mathbf{k} \parallel \mathbf{z}$), which is normal to the sample plane. The Faraday geometry corresponds to a magnetic field applied along the optical axis ($\mathbf{B}_\text{F} \parallel \mathbf{z}$), whereas in Voigt geometry, the field is applied in-plane along the $x$-axis ($\mathbf{B}_\text{V} \parallel \mathbf{x}$). A coordinate system is included for clarity. b) Energy level diagram and optical transitions for the 1$s$ exciton in Faraday (left) and Voigt (right) geometries. Optical transitions between the crystal ground state $|{\rm G}\rangle$ and bright exciton states (green) are shown with red/blue arrows for light propagating along the $z$-axis. The dark state $\ket{0,0}$ is separated from the bright states $\ket{1,(0,\pm1)}$ (the indices reflect the direction of the dipole moment) by an energy splitting $\Delta_{\rm X}$. In an external magnetic field, the exciton states undergo splitting and mixing. In Faraday geometry (left), mixed states remain inaccessible, and only $\ket{1,(\pm1)}$ are addressed optically with $\sigma^{\pm}$ polarized light, respectively. In Voigt geometry (right), the mixed states become accessible via linearly polarized light. In strong magnetic field, the inner and outer doublets are polarized perpendicular ($V$ polarization) and parallel ($H$ polarization) to the magnetic field direction $\mathbf{B}_\text{V}$, respectively. The level arrangement assumes $g_\text{e} > 0$ and $g_\text{h} < 0$, following Ref.~\citenum{24KoptevaAdvSci}, to provide a consistent representation of the observed splitting behavior of the exciton states.} 
    \label{fig:X_levels}
\end{figure*}

We consider band-edge 1$s$ exciton, which comprises conduction band electron with angular momentum $J_{\rm e}=1/2$ and valence band hole with the spin $J_{\rm h}=1/2$ at the R-point of the Brillouin zone. The corresponding optical transition has the photon energy $\hbar\omega_{\rm X}\approx1.51$~eV at crygenic temperature of $T=2$~K. More details on the electronic band structure of the bulk \mixperov{} perovskite cubic single crystal are reported in Refs.~\citenum{24KoptevaAdvSci, 22KirsteinAdvMat, 24KoptevaAdvSci_weak}. 

The energy structure comprises four states which are determined by the total angular momentum $J_{\rm X} = 0,1$ and its projection $J_{\rm X,z}$. Electron-hole exchange interaction lifts the degeneracy between bright triplet ($J_{\rm X}=1$) and dark singlet ($J_{\rm X}=0$) exciton levels by an energy $\BDSS{}$ as shown in Figure~\hyperref[fig:X_levels]{\ref*{fig:X_levels}b}. Assuming a cubic crystal lattice structure, no splitting of triplet states in zero magnetic field is expected. Each of these three states is characterized by an angular momentum projection $J_{\rm X,z}=0,\pm1$ along the $z$-axis set by the optical excitation direction $\mathbf{k}$. All states are labeled using $\ket{J_{\rm X},J_{\rm X,z}}$ notation in Figure~\hyperref[fig:X_levels]{\ref*{fig:X_levels}b}. Circularly polarized light $\sigma^{\pm}$ couples to the bright exciton states $\ket{1,\pm1}$, since for these states the electric dipole moments lies in the $xy$-plane. In contrast, the bright exciton state $\ket{1,0}$ has dipole moment parallel to the $z$-axis, making it inaccessible for light propagating along this direction. 

In external magnetic field $\mathbf{B}$, Zeeman splitting leads to lifting of triplet degeneracy. As result, all four states acquire different energy. The energy splitting between the $\ket{1,\pm1}$ states grows linearly with magnetic field: $\hbar\Omega_{+}=(g_\mathrm{e}+g_\mathrm{h}) \mu_B B$, where $\mu_B$ is the Bohr magneton and $g_{\rm e}$ and $g_{\rm h}$ are the electron and hole $g$-factors, respectively.  
In Figure~\hyperref[fig:X_levels]{\ref*{fig:X_levels}b} this corresponds to the inner doublet with energy splitting $\hbar\Omega_{\rm I}=\hbar\Omega_{+}$ assuming the $g$-factors of electron and hole have different signs, i.e.  $g_\mathrm{e}>0$ and $g_\mathrm{h}<0$\fcite{22KirsteinNatCom,22KirsteinAdvMat}. The other two states are mixed by magnetic field which leads to redistribution of oscillator strength with polarization parallel to magnetic field direction making the dark state optically active in Voigt geometry (see green arrows in Figure~\hyperref[fig:X_levels]{\ref*{fig:X_levels}b}). The energy splitting between the outer states $\hbar\Omega_{\rm O}=\hbar\sqrt{\Omega_{-}^2+\Omega_{\rm X}^2}$, where $\hbar\Omega_{\rm X}=\Delta_{\rm X}$ and  $\hbar\Omega_{-}=(g_\mathrm{e}-g_\mathrm{h}) \mu_B B$. 

The possibility to optically address different exciton states and their polarization depends on the direction of optical excitation ($\mathbf{k}\|\mathbf{z}$-axis) with respect to magnetic field. In Faraday geometry ($\textbf{B}_\text{F}\parallel\textbf{k}$), only the inner doublet is accessible (see Figure~\hyperref[fig:X_levels]{\ref*{fig:X_levels}a}).  In Voigt geometry ($\mathbf{B}_\text{V}\perp\mathbf{k}$), all optical transitions are allowed with inner doublet being polarized perpendicular to the magnetic field ($V$ polarization), and the outer doublet parallel to $\mathbf{B}$ ($H$ polarization). Explicit treatment of exciton splitting and selection rules for optical transitions is presented in Section~\ref{SI:Theory}, Supplementary Information. 
  
When a resonant femtosecond pulse spectrally covers several exciton transitions, it creates a coherent superposition of the corresponding states, resulting in quantum beats in the exciton emission.\fcite{bookScully}. Therefore, if the optical coherence of excitons is sufficiently long, one can directly measure $\hbar\Omega_{\rm I}$ and $\hbar\Omega_{\rm O}$. By analyzing their dependence on magnetic field, it becomes possible to evaluate $\Delta_{\rm X}$. Moreover, the contrast of the quantum beats in the outer doublet is determined by the brightening of the dark state, which is proportional to $(\Omega_{-}/\Omega_{\rm O})^2$.

\subsection{Photon echo in zero magnetic field}
\label{sec:exp}
The studied sample is a solution-grown single crystal of the composition \mixperov, featuring distinct facets and a size of several millimeters\fcite{17NazarenkoNPGAM,23GrisardNano}. At cryogenic temperatures, excitons are assumed to be weakly localized by composition fluctuations\fcite{23GrisardNano,24GrisardACS}. We emphasize that the localization potential is shallow ($\sim 16~$meV) and its size is in the range from 10 to 100~nm, which is larger than the exciton Bohr radius of about $\SI{5}{\nano\meter}$\fcite{20BaranowskiAEM, 20TamaratNatCom}. Therefore, weak confinement should not influence the exciton fine structure significantly and the bulk approximation is still valid. Optical spectroscopy was performed on the sample immersed in superfluid helium and kept at a temperature of $T=2$~K in a split-coil cryostat, with magnetic fields applied up to 6~T. 

\begin{figure}
    \centering
    \includegraphics[width=1\linewidth]{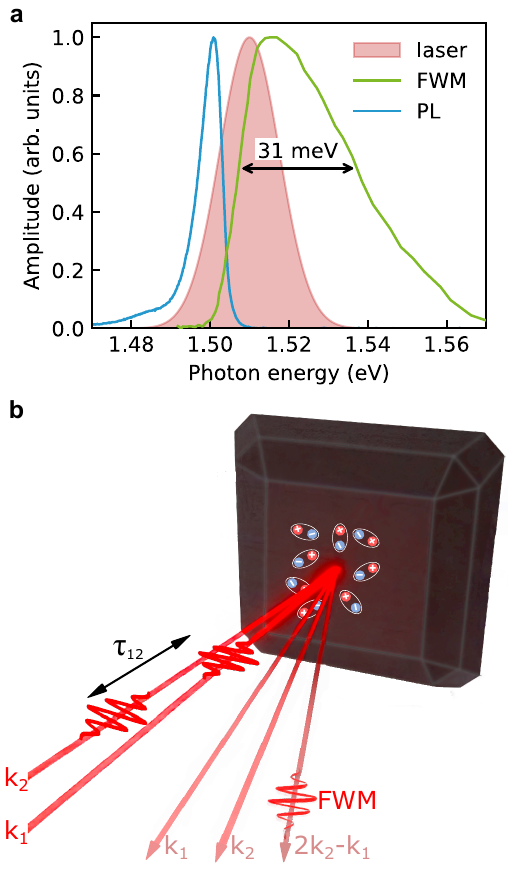}
    \caption{a) Four-wave-mixing (FWM, green) and photoluminescence (PL, blue) spectra of bulk mixed halide perovskite \mixperov{} sample, as previously reported in Ref.~\citenum{23GrisardNano}. The PL spectrum is measured under excitation with photon energy of 2.33~eV. The FWM spectrum was recorded using ps laser pulses. The excitation laser spectrum (shaded area) used in the PE experiments has a full width at half maximum of 18~meV and is centered at 1.51~eV. b) Schematic representation of the non-collinear transient FWM arrangement used for PE studies.}
    \label{fig:PE_schema}
\end{figure}

We use two-pulse photon echo spectroscopy to resolve quantum beats from the ensemble of excitons, which is otherwise hidden by the inhomogeneous broadening of optical transitions. Previous studies of photon echo in \mixperov{} crystals addressed exciton coherence in zero magnetic field using a sequence of spectrally narrow ($\sim 0.4$~meV) picosecond pulses\fcite{23GrisardNano,24GrisardACS}. This approach enabled spectrally resolved measurements by tuning the photon energy of the laser, yielding the spectral dependence of the four-wave-mixing (FWM) amplitude defined by $\chi^{(3)}$ and roughly following the absorption profile of the exciton resonance, which is shown in Figure~\hyperref[fig:PE_schema]{\ref*{fig:PE_schema}a } by the green line. It is centered at the photon energy of 1.514~eV with a full-width half maximum of 31~meV. The low-energy side is attributed to a ground state exciton emission with an inhomogeneous broadening of 16~meV, while the broad high-energy tail likely arises from excited exciton states. The photoluminescence spectrum is shown as the blue line in Figure~\hyperref[fig:PE_schema]{\ref*{fig:PE_schema}a}. Its maximum is shifted to lower energies by 14~meV relative to the FWM spectrum and is attributed to the recombination of long-lived, spatially separated electrons and holes\fcite{24KoptevaAdvSci}.

In this study, the energy splitting of exciton levels in a magnetic field may reach several meV, which does not allow all exciton states to be addressed when using picosecond pulses. 
Therefore, we perform photon echo experiments using spectrally broad laser pulses with a bandwidth of \SI{18}{\milli\electronvolt} and a duration of about $\SI{100}{\femto\second}$. The laser is tuned to a photon energy of 1.51~eV, i.e., 4~meV below the maximum of the FWM spectrum, to address low-energy exciton states, as shown in Figure~\hyperref[fig:PE_schema]{\ref{fig:PE_schema}a}. 
A sequence of two non-collinear pulses with wavevectors $\mathbf{k}_1$ and $\mathbf{k}_2$ at incidence angles of $\SI{4}{\degree}$ and $\SI{3}{\degree}$ is focused onto a spot of $\approx\SI{150}{\micro\meter}$. The pulses are delayed with respect to each other by $\tau_{12}$, as illustrated in Figure~\hyperref[fig:PE_schema]{\ref*{fig:PE_schema}b}. Due to the inhomogeneous broadening of exciton transitions, the resulting FWM signal is represented by a photon echo pulse delayed by $2\tau_{12}$ relative to the 1st pulse. The photon echo amplitude as a function of $\tau_{12}$ is measured in reflection geometry in the direction $\mathbf{k_{\rm FWM}} = 2\mathbf{k}_2 - \mathbf{k}_1$ using heterodyne detection\fcite{18PoltavstevPSS} (see Methods for details). All experiments are performed at low pulse fluences, kept below $\SI{2}{\micro\joule\per\centi\meter\squared}$, in order to remain in the $\chi^{(3)}$ regime. Under these conditions, the photon echo intensity depends linearly and quadratically on the electric field amplitudes of the 1st and 2nd pulses, respectively.

\begin{figure}
    \centering
    \includegraphics[width=\linewidth]{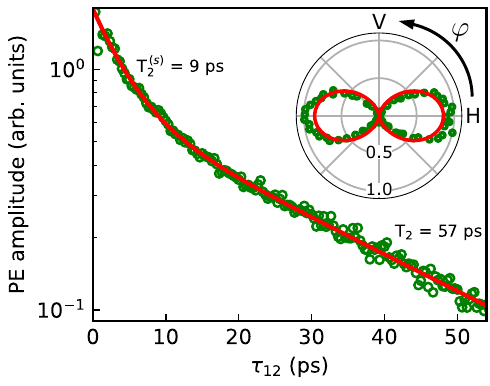}
    \caption{Decay of the photon echo (PE) amplitude at zero magnetic field ($B=0$) measured in lineraly co-polarized configuration. Solid red line is a fit with Equation~\eqref{eq:expexp}. Inset: Polar plot of PE amplitude as a function of angle $\theta$ between linear polarizations of the 1st and 2nd  pulse measured at $\tau_{12}=4$~ps. Solid red line is a fit with $\cos^2{\theta}$ which corresponds to typical exciton behavior\fcite{19PoltavtsevSR}. }
    \label{fig:0T}
\end{figure}
\noindent

The decay of the PE amplitude with increasing delay $\tau_{12}$ at zero magnetic field is presented in Figure~\ref{fig:0T}. It is well described by a double exponential decay
\begin{equation}
    P_0(\tau_{12}) = A^{(s)}\exp\biggl(-\frac{2\tau_{12}}{T_2^{(s)}}\biggr) + A \exp\biggl(-\frac{2\tau_{12}}{T_2}\biggr),
    \label{eq:expexp}
\end{equation}
where $A^{(s)}$ and $A$ denote the amplitudes of the short- and long-lived components, respectively. This yields decay times of $T_2^{(s)} = 9$~ps and $T_2 = 57$~ps. The shorter time is tentatively related to excitation of higher energy states with coherence times shortened by the energy relaxation to lower energy states. This is expected for a spectrally broad excitation laser and is consistent with the previous studies in Ref.~\cite{23GrisardNano}.    
Photon echo polarimetry, where the linear polarization of the excitation pulses as well as the PE detection is varied, is a powerful tool for distinguishing different resonantly excited complexes (e.g., neutral or charged excitons, biexcitons, etc.)\fcite{19PoltavtsevSR,22ArturACS}. Such measurements, in which the linear polarization of the 1st pulse and the PE detection are co-polarized while the polarization of the 2nd pulse is rotated, are shown in the inset of Figure~\ref{fig:0T}. In agreement with previous studies, we observe a two-leaf polar pattern characteristic of excitons\fcite{24GrisardACS}. 

\subsection{Quantum beats and evaluation of bright-dark exciton splitting}
\label{sec:result}
\begin{figure*}
    \centering
    \includegraphics[width=\linewidth]{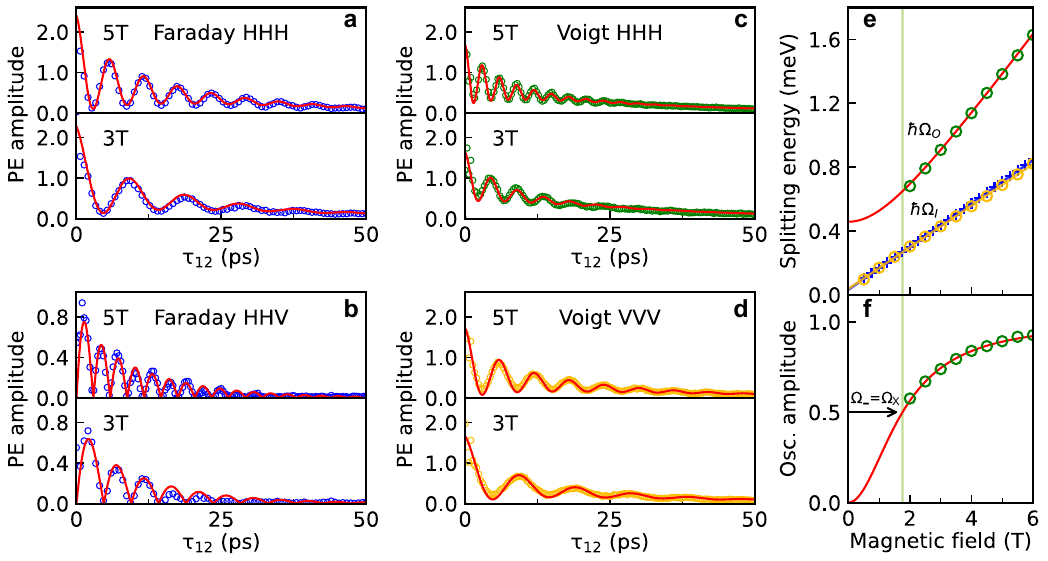}
    \caption{Photon echo oscillations in magnetic field $B$ for different polarization configurations. a)-d): Time-resolved photon echo amplitude (arb. units) for $\SI{3}{\tesla}$ and $\SI{5}{\tesla}$. Each legend indicates the horizontal ($H\parallel\mathbf{x}$) and vertical ($V\parallel\mathbf{y}$) polarization of the first and second excitation pulse and detection. Solid curves represent the results of the modeling described by Equations~\eqref{eq:PEeqs_F_HHH}-\eqref{eq:PEeqs_V_HHH}. Full set of data and the details of analysis procedure can be found in the Supplementary Information~\ref{SI-fit}. The evaluated parameters are summarized in Table~\ref{tab:fit}. e) Magnetic field dependence of the energy splittings between the inner $\hbar\Omega_{\rm I}$ and outer $\hbar\Omega_{\rm O}$ excitonic doublets. Plus and circle symbols correspond to the data obtained in Faraday and Voigt geometries, respectively. Solid blue/yellow line is a linear fit with $\hbar\Omega_{\rm I}+\delta$ with $\delta\approx35~\mu$eV. Red line represents the fit with $\hbar\Omega_{\rm O} = \hbar\sqrt{\Omega_{-}^2+\Omega_{\rm X}^2}$. e) Solid red line is the fit with $\Omega^2_{-}/(\Omega_{-}+\Omega_{\rm X})^2$ where $\Delta_{\rm X}=\hbar\Omega_{\rm X}=0.46$~meV. The arrow marks the point where the Zeeman splitting $\hbar\Omega_{-}$ matches the bright-dark splitting $\Delta_{\rm X}$. f) Magnetic field dependence of the oscillations amplitude $C=(\Omega_{-}/\Omega_{\rm O})^2$. The circles are extracted from the fit of $HHH$ transients in Voigt geometry using Equation~\eqref{eq:PEeqs_V_HHH} with fixed value of $g_\text{e}-g_\text{h}=4.52$ which is evaluated from the fit of $\hbar\Omega_{\rm O}(B)$ in panel.}
    \label{fig:3and5T}
\end{figure*}

External magnetic field lifts the degeneracy of bright triplet states and mixes the dark state with the bright exciton  linearly polarized along $\mathbf{B}$, as discussed in Section~\ref{sec:theory}. Therefore, quantum beats in the intensity and polarization of the PE are expected. Here, we use linearly polarized excitation and detection. We use a three letter ($ABC$)-notation to indicate the polarization configuration where $A$ and $B$ correspond to polarization of excitation while $C$ stands for detection being set to $H\|x$ or $V\|y$ directions. Excitation pulses are co-polarized, while detection can be adjusted to the same or crossed polarization. Other polarization configurations are discussed in Section~\ref{SI-pola}, Supplementary Information. 
Figure~\hyperref[fig:3and5T]{\ref*{fig:3and5T}a-d} summarizes PE decay in magnetic fields of 3 and 5~T for different polarization configurations and magnetic field geometries. As expected, we observe oscillations of the PE amplitude, with a period that decreases as the magnetic field increases from 3 to 5~T. 

\begin{table*}
    \centering
\caption{The exciton parameters obtained by applying the model described by Equations~\eqref{eq:PEeqs_F_HHH}-\eqref{eq:PEeqs_V_HHH} and the fitting procedure outlined in Supplementary Information~\ref{SI-fit}. The reported uncertainties reflect only the statistical errors of the fitting routine and do not account for sample inhomogeneities, which lead to parameter fluctuations of up to 5\% across different regions of the sample.}
    \begin{tblr}{colspec={c c c c c c}}
        \hline
        {B-field geometry} & \SetCell[c=2]{c}{$g$-factors} && {splitting parameter} & \SetCell[c=2]{c}{QB decay times} \\
        & {$g_\mathrm{e}-g_\mathrm{h}$} & {$g_\mathrm{e}+g_\mathrm{h}$} & {$\Delta_{\text{X}}$ (meV)} & {$T_\text{I}^\text{QB}$ (ps)} & {$T_\text{O}^\text{QB}$ (ps)} \\
        \hline
        {Voigt}   & {4.52$\pm$0.01} & {2.24$\pm$0.02} & {0.46$\pm$0.01} & {47$\pm$0.5} & {21$\pm$0.4}  \\
        {Faraday} & {}     & {2.35$\pm$0.02} & {} & {47$\pm$0.5} & {}      \\
        \hline
    \end{tblr}
    \label{tab:fit}
\end{table*}

In Faraday geometry, the signal does not depend on the polarization direction of the excitation pulses. In a magnetic field, the PE signal also appears for crossed detection polarization ($HHV$), as shown in Figure~\hyperref[fig:3and5T]{\ref*{fig:3and5T}b}, which is expected from the fact that optical transitions remain circularly polarized (see Figure~\hyperref[fig:X_levels]{\ref*{fig:X_levels}a-d}). In Voigt geometry we use $HHH$ and $VVV$ configurations being directed parallel and perpendicular to magnetic field direction, respectively. In this case we detect PE signal in the same polarization as excitation pulses only. Note that in the Voigt geometry we observe pure modulation of the PE intensity, whereas in the Faraday geometry an additional rotation of the PE polarization occurs. We also observe that for the same magnetic field the oscillation period for $HHH$ polarization is smaller than in $VVV$ polarization. 

The solution of Lindblad equation (see Section~\ref{SI:Theory}, Supplementary Information) yields the following expression for the PE electric-field amplitude in the Faraday geometry for $HH$-polarized excitation sequence
\begin{align}
    \mathbf{P_{\rm F}^\textit{HH}} = &
        P_0(\tau_{12})\left[
             \mathbf{e}_x \left( 
                1 + \cos(\Omega_{\rm I}\tau_{12})
                \exp\left(-\frac{2\tau_{12}}{T^{\rm QB}_{\rm I}}\right)
            \right)
        \right.
        \nonumber
    \\
        & \hspace{1cm} \left.
        +\  \mathbf{e}_y\,\sin(\Omega_{\rm I}\tau_{12})\exp\left(-\frac{2\tau_{12}}{T^{\rm QB}_{\rm I}}\right)
        \right],
    \label{eq:PEeqs_F_HHH}
\end{align}
and in the Voigt geometry for $VV$ and $HH$ excitation
\begin{align}
    \mathbf{P_{\rm V}^\textit{VV}} &= P_0(\tau_{12})\mathbf{e}_y \left[ 1 + \cos(\Omega_{\rm I} \tau_{12})\exp\left(-\frac{2\tau_{12}}{T^{\rm QB}_{\rm I}}\right) \right], 
    \label{eq:PEeqs_V_VVV} 
\end{align}
\begin{align}
\mathbf{P_{\rm V}^\textit{HH}} 
    &= P_0(\tau_{12})\mathbf{e}_x
       \left[
           1 + \frac{\Omega_{\rm X}^{2}}{\Omega_{\rm O}^{2}}            
       \right. \nonumber
\\
    &\left.\qquad
       + \frac{\Omega_{-}^{2}}{\Omega_{\rm O}^{2}} \cos\!\left(\Omega_{\rm O} \tau_{12} \right)
       \exp\left(-\frac{2\tau_{12}}{T^{\rm QB}_{\rm O}}\right)
       \right].
\label{eq:PEeqs_V_HHH}
\end{align}
Here $\mathbf{e}_x$ and $\mathbf{e}_y$ are the unit polarization vectors along $x$- and $y$-axes corresponding to $H$ and $V$ linear polarizations, respectively.  An additional decay of oscillating components with time constants $T^{\rm QB}_{\rm I}$ and $T^{\rm QB}_{\rm O}$ is due to dephasing defined by the fluctuations of exciton fine structure splittings in the ensemble of excitons. 

Using Equations \eqref{eq:PEeqs_F_HHH}-\eqref{eq:PEeqs_V_HHH} we evaluate the main parameters responsible for energy splitting: $g_\mathrm{e}$, $g_\mathrm{h}$, $\Delta_{\rm X}$, coherence times $T_2^{(s)}$, $T_2$, and dephasing times of quantum beats $T^{\rm QB}_{\rm I}$ and $T^{\rm QB}_{\rm O}$. The results are summarized in Table~\ref{tab:fit}. First, the frequencies $\Omega_{\rm I}$ (from  $HHH$, $HHV$ in Faraday and $VVV$ in Voigt fields) and $\Omega_{\rm O}$ (from $HHH$ signal in Voigt field) are evaluated. The magnetic field dependence of splittings for different configurations are presented in Figure~\hyperref[fig:3and5T]{\ref*{fig:3and5T}e}. The inner doublet $\hbar\Omega_{\rm I}$ follows linear dependence (green data points) with the slope given by $g_\mathrm{e}+g_\mathrm{h} \approx 2.3$ and small offset of about $\delta=31-38~\mu$eV. The $HHH$ data in Voigt geometry (gray circles) do not allow precise evaluation of the oscillations at $B < \SI{2}{\tesla}$ due to the low oscillation amplitude and are therefore not analyzed in this magnetic-field range. For larger $B$, the dependence of the outer doublet splitting $\hbar\Omega_\text{O}$ appears nearly linear, and the expected $\sqrt{(\hbar\Omega_-)^2 + \BDSS^2}$ dependence is not clearly resolved. However, the presence of a significant offset, visible even at high magnetic fields, provides clear evidence for the bright-dark exciton splitting. From the fit, we extract $g_\text{e} - g_\text{h} = 4.52$ and $\BDSS = \SI{0.46}{\milli\electronvolt}$.

Since $\BDSS{}$ affects the oscillation amplitude in the $HHH$ configuration, its magnitude can be independently verified via the magnetic-field dependence of the oscillation strength. Figure~\hyperref[fig:3and5T]{\ref*{fig:3and5T}f} shows the extracted oscillation amplitude $C(B)$ as a function of applied magnetic field. From a fit with $\Omega^2_{-}/(\Omega_{-}+\Omega_{\rm X})^2$, we obtain $\BDSS{} = \SI{0.46 \pm 0.01}{\milli\electronvolt}$, in good agreement with the value derived from the analysis based on energy splitting in Figure~\hyperref[fig:3and5T]{\ref*{fig:3and5T}e}. This result supports the consistency of the extracted exchange splitting across both fitting approaches. Another feature related to the oscillation contrast is its decrease with increasing $\tau_{12}$.  This indicates that the quantum beats decay on a shorter time scale compared to the non-oscillating part of the signal being governed by the quantum beats dephasing time. This behavior is especially pronounced in the $HHH$ Voigt configuration, where $T^{\rm QB}_{\rm O}=21$~ps is more than twice shorter than $T^{\rm QB}_{\rm I}=47$~ps. More details on the fitting procedure are provided in Supplementary Information~\ref{SI-fit}.

\subsection{Discussion}
\label{sec:Discussion}

Using the parameters of Table~\ref{tab:fit}, we obtain the Land\'e $g$-factors for electrons, $g_{\mathrm{e}} = 3.38$, and holes, $g_{\mathrm{h}} = -1.14$. These values are in good agreement with those reported for perovskite crystals of the same composition, as obtained from time-resolved pump-probe Kerr-rotation\fcite{22KirsteinNatCom,22KirsteinAdvMat} and optical orientation\fcite{24KoptevaAdvSci}. The magnitude of the exciton bright-dark splitting $\Delta_{\rm X}=0.46$~meV is also consistent with previous estimations of this quantity for bulk FAPbI$_3$ (0.55~meV) and CsPbI$_3$ (0.75~meV)  reported by Tamarat \textit{et al.}\fcite{20TamaratNatCom}.

Notably, the dephasing of the quantum beats is not negligible. Here, both $T^\text{QB}_{\rm I}$ and $T^\text{QB}_{\rm O}$ remain insensitive to variations of the applied magnetic field, allowing us to exclude the spread of the $g$-factors as the origin of the dephasing.  
Nuclear fluctuations, corresponding to effective magnetic fields in the order of 10~mT, have been observed in the same materials\fcite{24KudlacikACS}. In this case the expected energy splitting of exciton corresponds to few $\mu$eV, which is too small compared to the homogeneous linewidth of the exciton transition $2\hbar/T_2=23~\mu$eV as follows from Figure~\hyperref[fig:0T]{\ref*{fig:0T}}. In the case of quantum beats between the outer exciton states, the dephasing can be attributed to a spread in $\Delta_{\rm X}$, which we estimate as $2\hbar/T^{\rm QB}_{\rm O} = 60~\mu\text{eV}$.

Interestingly, the dephasing of oscillations is also observed for the quantum beats between the inner exciton states, with $T^{\rm QB}_{\rm I} = 47$~ps. This indicates an additional splitting of the bright exciton states due to the anisotropic electron-hole exchange interaction\fcite{Goupalov98,Goupalov98b,18NestoklonPRB}. In this case, the triplet degeneracy is lifted at low temperatures because of the low structural symmetry of the perovskite crystal\fcite{19BaranowskiNanoLet} or localization potential anisotropy with a random distribution, as the excitons are localized on a fluctuating landscape\fcite{23GrisardNano}. For such ensemble of excitons, we expect zero splitting on average, with a dispersion given by $2\hbar/T^{\rm QB}_{\rm I} = 28~\mu\text{eV}$. This also explains the deviations in modeling the PE transients at low magnetic fields below 2~T, as discussed in detail in Supplementary Section~\ref{SI-fit}. The presence of zero-field splitting of bright excitons is further supported by the observation of a small offset, $\delta \sim 35~\mu\text{eV}$ (see Figure ~\hyperref[fig:3and5T]{\ref*{fig:3and5T}e}), which is in very good agreement with the dispersion estimated from $T^{\rm QB}_{\rm I}$.

\section{Conclusion}

In conclusion, we demonstrate that polarization-sensitive photon echo spectroscopy in magnetic fields provides a powerful tool to uncover the energy structure of excitons in lead halide perovskites, which is otherwise hidden by inhomogeneous broadening. The observation of exciton quantum beats in both Faraday and Voigt geometries offers a precise probe of the energy splittings among the four 1$s$ excitonic states and allows determination of the fine-structure splitting. In this work, we apply this technique to evaluate the bright-dark splitting in mixed-halide perovskite \mixperov{} crystals. We evaluate the $g$-factors of electrons $g_\mathrm{e}=3.38$ and holes $g_\mathrm{h}=-1.14$ with bright-dark splitting $\Delta_{\rm X}=0.46$~meV. The latter has a dispersion of about 0.06~meV arising from localization of excitons on a random potential. The dephasing of the bright exciton quantum beats is attributed to a weak splitting of the bright-state on the order of $0.035$~meV, which possibly result from the anisotropy of the localization potential with random distribution. This is consistent with the expectation that the bright-exciton fine-structure splitting is smaller than the bright-dark splitting\fcite{02BayerPRB}. The quantum beats between the exciton states persist on timescales comparable to the exciton lifetime  revealing remarkably robust spin and optical coherences.

The described approach for investigating the fine structure in an ensemble of excitons can also be adapted to other types of localized excitons, such as those in perovskite nanostructures (e.g., two- and zero-dimensional systems). In contrast to conventional pump-probe Faraday/Kerr rotation, this method enables the observation of level splittings even in longitudinal magnetic fields and is additionally sensitive to crystal anisotropy\fcite{21TrifonovPRB}. Therefore, it opens new possibilities for exploring exciton fine structure in a wide range of low-dimensional perovskite-based nanostructures.

\section{Experimental Section}
The \mixperov{} crystals were grown using a modified version of the method previously employed to synthesize formamidinium iodide (FAI) crystals (Ref.~\citenum{17NazarenkoNPGAM}). Details of the crystal preparation are provided in Ref.~\citenum{23GrisardNano}.

Photon echo spectroscopy was performed using transient four-wave mixing with heterodyne detection. The sample was excited by a sequence of two pulses generated by a tunable Ti:Sa mode-locked oscillator operating at a repetition rate of 75.7~MHz, with a pulse duration of approximately 100~fs. The sample was mounted in a split-coil cryostat at a temperature of 2~K, immersed in superfluid liquid helium. To prevent mechanical stress during cooling, the mixed-halide perovskite sample was held in a paper envelope. Magnetic fields up to 6~T were applied in both Faraday and Voigt geometries.

The four-wave mixing signal was detected in reflection geometry along the phase-matched direction $2\mathbf{k}_2-\mathbf{k}_1$, where $\mathbf{k}_1$ and $\mathbf{k}_2$ are the wavevectors of excitation pulses 1 and 2, incident on the sample at angles of 4$^\circ$ and 3$^\circ$, respectively. The frequency of the first beam was shifted using an acousto-optical modulator by $-81$~MHz. The frequency of additional reference pulse was shifted by $+80$~MHz. The cross-correlation signal between the four-wave mixing signal and a reference pulse with delay $\tau_{\rm ref}$ was detected with a silicon photodiode. Heterodyne detection was performed at the difference frequency 1~MHz using a lock-in amplifier, while the second excitation beam was mechanically chopped at approximately 800~Hz to suppress scattered light, resulting in a double lock-in detection scheme.

Scanning $\tau_{\rm ref}$ allows temporal resolution of the four-wave mixing signal\fcite{18PoltavstevPSS}. This scan yields a photon echo pulse centered at $\tau_{\rm ref} = 2\tau_{12}$. The photon echo duration is approximately twice that of the excitation pulse, consistent with the regime in which the inhomogeneous broadening is comparable to the spectral bandwidth of the pulse.

\section*{Acknowledgment} 
We are grateful to N.E. Kopteva for useful discussions. The Dortmund group acknowledge financial support from the Deutsche Forschungsgemeinschaft within the SPP 2196 (project no. 506623857). The work at ETH Z\"urich (O.H., D.N.D., and M.V.K.) was financially supported by the Swiss National Science Foundation (grant agreement 200020E 217589, funded through the DFG-SNSF bilateral program) and by ETH Z\"urich through ETH+ Project SynMatLab.

\section*{Conflict of Interest} 
The authors declare that they have no competing interests.\\[0.5cm]
\textbf{ORCID}\\
M. Alex Hollberg:      0000-0003-0982-9213 \\
Artur~V.~Trifonov:   0000-0002-3830-6035 \\
Stefan Grisard:      0000-0002-5011-3455 \\
Mikhail.~O.~Nestoklon:     0000-0002-0454-342X \\
Oleh Hordiichuk:     0000-0001-7679-4423 \\
Dmitry N. Dirin:     0000-0002-5187-4555  \\
Maksym V. Kovalenko: 0000-0002-6396-8938  \\
Dmitri R. Yakovlev:  0000-0001-7349-2745  \\
Manfred Bayer:       0000-0002-0893-5949  \\
Ilya A.~Akimov:      0000-0002-5011-3455  \\

\bibliography{lit}

\appendix
\onecolumngrid

\newpage

\setcounter{figure}{0}
\renewcommand{\thefigure}{S\arabic{figure}}
\renewcommand\theHfigure{Sfigure.\arabic{figure}}  

\section{Supporting information}

\subsection{Theory of PE quantum beats from excitons in magnetic field}
\label{SI:Theory}

The signal of the photon echo is proportional to the corresponding components of polarization vector at the time $\tau_{12}$ after second pulse, where $\tau_{12}$ is the time between first and second pulse\fcite{bookBerman}. The signal is calculated by tracing the components of the density matrix contributing to the photon echo signal. Below we neglect fine structure splitting of the bright exciton state and assume that the pulse duration is small as compared with any characteristic times in the system.

\begin{figure}[H]
    \centering
\tdplotsetmaincoords{60}{140}
\begin{tikzpicture}[tdplot_main_coords, scale=2, >=latex']
  \draw[tdplot_screen_coords,color=blue,opacity=0] (-0.4,-0.7) rectangle (1.3,1.1);
  \node[below right,tdplot_screen_coords] at (-0.7,1.3) {a)};
    \tdplotsetrotatedcoords{0}{-90}{0}
    \draw[thick,tdplot_rotated_coords,->] (0,0,0) -- (1.0,0,0) node[anchor=south]{$x'$};
    \draw[thick,tdplot_rotated_coords,->] (0,0,0) -- (0,1.0,0) node[anchor=west]{$y'$};
    \draw[thick,tdplot_rotated_coords,->] (0,0,0) -- (0,0,1.0) node[anchor=south]{$z'$};

    \draw[line width=2.0,tdplot_rotated_coords,->] (0.01,0.2,1.5) -- (0.01,0.2,0.3) node[anchor=north west]{${\bf B}$};

    \coordinate[tdplot_rotated_coords] (Shift) at (-0.1,-0.6,-0.8);
    \tdplotsetrotatedcoords{90}{-90}{0}
    \tdplotsetrotatedcoordsorigin{(Shift)}
    \draw[thick,tdplot_rotated_coords,->] (0,0,0) -- (0.5,0,0) node[anchor=south]{$x$};
    \draw[line width=1.5,tdplot_rotated_coords,->] (0,0,0) -- (0,0.5,0) node[anchor=west]{${\bf k}$};
    \draw[thick,tdplot_rotated_coords,->] (0,0,0) -- (0,0,0.5) node[anchor=south]{$y$};
\end{tikzpicture}
\begin{tikzpicture}[tdplot_main_coords, scale=2, >=latex']
  \draw[tdplot_screen_coords,color=blue,opacity=0] (-0.1,-0.7) rectangle (1.9,1.1);
  \node[below right,tdplot_screen_coords] at (-0.4,1.3) {b)};
    \tdplotsetrotatedcoords{0}{-90}{0}
    \draw[thick,tdplot_rotated_coords,->] (0,0,0) -- (1.0,0,0) node[anchor=south]{$x'$};
    \draw[thick,tdplot_rotated_coords,->] (0,0,0) -- (0,1.0,0) node[anchor=west]{$y'$};
    \draw[thick,tdplot_rotated_coords,->] (0,0,0) -- (0,0,1.0) node[anchor=south]{$z'$};

    \draw[line width=2.0,tdplot_rotated_coords,->] (0.01,0.2,1.5) -- (0.01,0.2,0.3) node[anchor=north west]{${\bf B}$};
    
    \coordinate[tdplot_rotated_coords] (Shift) at (-0.4,1.6,0.1);
    \tdplotsetrotatedcoordsorigin{(Shift)}
    \tdplotsetrotatedcoords{90}{-90}{0}
    \draw[thick,tdplot_rotated_coords,->] (0,0,0) -- (0.5,0,0) node[anchor=south]{$y$};
    \draw[thick,tdplot_rotated_coords,->] (0,0,0) -- (0,0.5,0) node[anchor=west]{$x$};
    \draw[line width=1.5,tdplot_rotated_coords,->] (0,0,0) -- (0,0,0.5) node[anchor=south]{${\bf k}$};
\end{tikzpicture}

    \caption{Scheme of coordinate systems aligned with magnetic field $x'y'z'$ and the coordinate system with respect to pump/probe pulse direction $xyz$.}
    \label{fig_cs_si}
\end{figure}
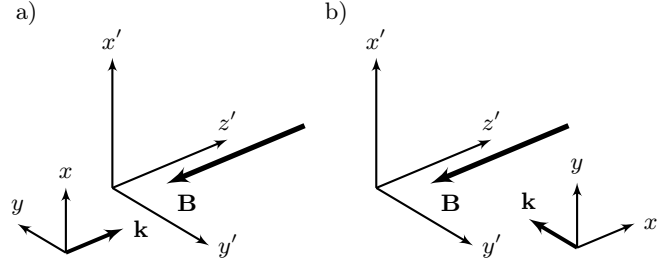

We work in the basis formed by the ground state $\ket{G}$, three linearly polarized bright exciton states $\ket{X}$, $\ket{Y}$, $\ket{Z}$ and the dark exciton state $\ket{D}$. The linearly polarized components are chosen to be aligned with magnetic field, with axis $z^\prime$ along magnetic field.  Note that in the main text the coordinate system is chosen with $z$ axis along excitation light beam $\mathbf{k}$ (see Figure~\hyperref[fig:X_levels]{\ref*{fig:X_levels}a}). For Faraday geometry the coordinate systems are equivalent but $y$ points in opposite direction, while in Voigt geometry it is necessary to exchange $z^\prime$ to $x$ and $x^\prime$ to $y$, see Figure~\ref{fig_cs_si}. The Hamiltoninan of exciton in this basis reads as 
\begin{equation}\label{eq:HBforPE}
  \mathcal{H}_B = \hbar
  \begin{blockarray}{rc ccc c}
    & \ket{G} & \ket{X} & \ket{Y} & \ket{Z} & \ket{D}\\
    \begin{block}{r(c|cc|cc)}
      \bra{G}\phantom{t}  & 0  & 0       & 0  & 0  & 0\\ \cline{2-6}
      \bra{X}\phantom{t}  & 0  & \omega   & -i\Omega_+/2  & 0  & 0 \\ 
      \bra{Y}\phantom{t}  & 0  & i\Omega_+/2 & \omega & 0  & 0 \\ \cline{2-6}
      \bra{Z}\phantom{t}  & 0  & 0  & 0  & \omega & \Omega_-/2 \\ 
      \bra{D}\phantom{t}  & 0  & 0  & 0  & \Omega_-/2  & \omega-\Omega_{\rm X} \\ 
    \end{block}
  \end{blockarray}\,,
\end{equation}
where $\omega$ is the bright exciton frequency, magnetic field splittings are $\hbar\Omega_+ = (g_\mathrm{e}+g_\mathrm{h})\mu_B B$, $\hbar\Omega_- = (g_\mathrm{e}-g_\mathrm{h})\mu_B B$, and the splitting between dark and bright exciton states is given by $\hbar\Omega_{\rm X} = \Delta_{\rm X}$. 
Below we will use also eigenfrequencies of the states with spin projection zero $\omega_{1,2} = \omega-\Omega_{\rm X}/2 \pm \Omega_{\mathrm{O}}/2$, where we defined the splitting for this pair of states $\Omega_{\mathrm{O}} = \sqrt{\Omega_{-}^2+\Omega_{\rm X}^2}$.

\paragraph{Dynamics of states.} In this section, we assume that the dynamics is described by Lindblad equation with the Hamiltonian \eqref{eq:HBforPE} and coherence time $T_2$. Then, the dynamics of the exciton may be calculated analytically. The dynamics for the density matrix components associated with pair of states $\ket{X}$, $\ket{Y}$ is 
\begin{equation}\label{eq:rhoGXY_B}
\begin{split}
  \rho_{GX} & = e^{i\omega t}\Gamma_0 \left( \phantom{-} \rho_{GX}^{0} \cos{\frac{\Omega_+ t}2} + \rho_{GY}^{0} \sin{\frac{\Omega_+ t}2} \right)\,,\\
  \rho_{GY} & = e^{i\omega t}\Gamma_0 \left( - \rho_{GX}^{0} \sin{\frac{\Omega_+ t}2} + \rho_{GY}^{0} \cos{\frac{\Omega_+ t}2}  \right)\,,\\
\end{split}
\end{equation}
and the complex conjugated equations for the pair $\rho_{XG}$, $\rho_{YG}$. With single decoherence time, the decay is given by $\Gamma_0 = \exp [-t/(2T_2)]$, this result will be later generalized to two-exponent function.

For other pair of states, the dynamics is a bit more complex: 
\begin{equation}\label{eq:rhoGZD_B}
\begin{split}
  \rho_{GZ} & = \Gamma_0 \left[ e^{i\omega_{1} t} \frac{\Omega_{\rm O}+\Omega_{\rm X}}{2\Omega_{\rm O}} + e^{i\omega_{2} t} \frac{\Omega_{\rm O}-\Omega_{\rm X}}{2{\Omega_{\rm O}}} \right] \rho_{GZ}^{0} +
                \Gamma_0 \left[ e^{i\omega_{1} t} - e^{i\omega_{2} t} \right] \frac{\Omega_{-}}{2\Omega_{\rm O}} \rho_{GD}^{0} \,,\\
  \rho_{GD} & = \Gamma_0 \left[ e^{i\omega_{1} t} \frac{\Omega_{\rm O}-\Omega_{\rm X}}{2\Omega_{\rm O}} + e^{i\omega_{2} t} \frac{\Omega_{\rm O}+\Omega_{\rm X}}{2\Omega_{\rm O}} \right] \rho_{GD}^{0} +
                \Gamma_0 \left[ e^{i\omega_{1} t} - e^{i\omega_{2} t} \right] \frac{\Omega_{-}}{2\Omega_{\rm O}} \rho_{GZ}^{0} \,,\\
\end{split}
\end{equation}
and complex conjugated for $\rho_{ZG}$, $\rho_{DG}$.

\paragraph{Pulse action.} Assuming the pulse duration is short as compared with all characteristic times in the system, its action on the exciton density matrix relevant for the photon echo calculations may be summarized as follows: 
\begin{equation}\label{eq:pulse1}
  \rho_{GX}^{1+} = \theta_{1x}\,,\;\;\; \rho_{GY}^{1+} = \theta_{1y}\,,\;\;\; \rho_{GZ}^{1+} = \theta_{1z}\,,\;\;\; \rho_{GD}^{1+} = 0\,,
\end{equation}
where $\theta_{ni}$ is the $n$-th pulse area in $i$-th polarization. $\rho^{1+}_{j}$ is the $j$-th component of density matrix after 1st pulse.
Density matrix components after 2nd pulse $\rho^{2+}_{j}$ are connected with components before the pulse $\rho^{2-}_{j}$ as: 
\begin{equation}\label{eq:pulse2}
\begin{split}
  \rho_{XG}^{2+} &= \theta_{2x}^2 \rho_{GX}^- + \theta_{2x}\theta_{2y} \rho_{GY}^- + \theta_{x}\theta_{z} \rho_{GZ}^- \,,\\ 
  \rho_{YG}^{2+} &= \theta_{2y}^2 \rho_{GY}^- + \theta_{2y}\theta_{2x} \rho_{GX}^- + \theta_{y}\theta_{z} \rho_{GZ}^- \,,\\ 
  \rho_{ZG}^{2+} &= \theta_{2z}^2 \rho_{GZ}^- + \theta_{2z}\theta_{2x} \rho_{GX}^- + \theta_{z}\theta_{y} \rho_{GY}^- \,,\\
  \rho_{DG}^{2+} &= 0\,.
\end{split}
\end{equation}

\paragraph{Photon echo: Faraday configuration.} In Faraday configuration, we are interested in the scheme when both pulses are linearly co-polarized. Electric field is normal to the magnetic field direction, i.e. lies in $x^\prime y^\prime$ plane. If fine structure splitting is neglected, the axis choice is arbitrary and without loss of generality we assume that $H$-polarization is directed along $x^\prime$ and $V$-polarization is along $y^\prime$. After first pulse, we have only one non-zero density matrix component relevant for the PE calculation $\rho_{GX}^{1+}=1$ (we omit pulse area which gives the same amplitude in all equations), which gives before 2nd pulse to
\begin{equation}
  \rho_{GX}^{2-} = e^{i\omega \tau_{12}}\Gamma_0 \cos{\frac{\Omega_+ t }2} \,,\;\;\;\;  \rho_{GY}^{2-} =  - e^{i\omega \tau_{12}} \Gamma_0 \sin{\frac{\Omega_+ t}2} \,.
\end{equation}
Only $\rho_{GX}^{2-}$ is transformed by the H pulse to the relevant components after 2nd pulse, that is $\rho_{XG}^{2+}=\rho_{GX}^{2-}$ and the same dynamics leads to the following result: 
\begin{equation}\label{eq:PFHH}
  {\bf P}^{HH}_{F} \propto {\bf e}_{x\prime}  \rho_{XG}^{PE}  + {\bf e}_{y\prime} \rho_{YG}^{PE} = \frac{\Gamma_0^2}2 \left[ 
    {\bf e}_{x\prime} \left( 1+\cos\Omega_+\tau_{12} \right) - 
    {\bf e}_{y\prime} \sin\Omega_+\tau_{12}
  \right]\,.
\end{equation}
Analogously, in case of second V pulse only non-zero component relevant for PE is $\rho_{YG}^{2+}=\rho_{GY}^{2-}$, this results in 
\begin{equation}\label{eq:PFHV}
  {\bf P}^{HV}_{F} \propto 
  \frac{\Gamma_0^2}2 \left[ 
    \left(\cos\Omega_+\tau_{12} -1 \right)  {\bf e}_{x\prime}  - 
    {\bf e}_{y\prime} \sin\Omega_+\tau_{12}
  \right]\,.
\end{equation}

\paragraph{Photon echo: Voigt configuration.} In the Voight configuration, the linear horizontal polarization $H$ drives the $\ket{Z}$ component of the bright exciton and the vertical $V$ polarization the $\ket{Y}$ component. Without fine-structure splitting, they are fully independent, dynamics converts only within $\ket{Z}$ and dark state $\ket{D}$, while $\ket{Y}$ is mixed by the magnetic field with the inactive in this geometry state $\ket{X}$. The calculations for the $V$ polarization fully reproduces $HHH$ result in the Faraday configuration, see above. However, the $x$-component (which was $y$ in $HHH$ geometry) should be dropped:
\begin{equation}
  {\bf P}^{VV}_{V} \propto \frac{\Gamma_0^2}2 {\bf e}_{y\prime} \left( 1+\cos\Omega_+\tau_{12} \right)\,.
\end{equation}

The calculations of the $HH$ result is different, we give it in more detail. First pulse creates $\rho_{GZ}^{1+}=1$, which evolves to 2nd pulse into ($\rho_{GD}$ is not relevant for the result)
\begin{equation}
  \rho_{GZ}^{2-} = \Gamma_0 \left( 
  e^{i\omega_1\tau_{12}} \frac{\Omega_{\rm O}+\Omega_{\rm X}}{\Omega_{\rm O}} +
  e^{i\omega_2\tau_{12}} \frac{\Omega_{\rm O}-\Omega_{\rm X}}{\Omega_{\rm O}}
  \right)\,,
\end{equation}
which is converted into $\rho_{ZG}^{2+}$ by the second pulse and evolves to the photon echo time into
\begin{equation}
  {\bf P}^{HH}_{V} {\propto} 2 \Gamma_0^2 {\bf e}_{z\prime} \left[ 
    \frac{\Omega_{\rm O}^2+\Omega_{\rm X}^2}{\Omega_{\rm O}^2} +
    \frac{\Omega_{\rm O}^2-\Omega_{\rm X}^2}{\Omega_{\rm O}^2}\cos(\Omega_{\rm O}\tau_{12})
  \right]
\end{equation}
Note that in the main text the coordinate system aligned with the wave vector and not with the magnetic field is used. Also, in the main text instead of $\Omega_+$, the notation $\Omega_I\equiv \Omega_+$ is used to stress that this is the splitting of ``inner'' doublet.

Possible fluctuations of $g$-factor would result in the modification of this result. However, the spread of $g$-factors leads to the shortening of the decoherence time as a function of magnetic field \fcite{YakovlevBayer}. Contrarily, $T_2^*$ is approximately constant as a function of magnetic field according to experimental data and we conclude that the $g$-factor is constant within the sample with a good precision. Though, in zero magnetic field the fine structure splitting of the bright exciton neglected above leads to the oscillations of the signal at the frequencies of splitting between the eigenstates of the bright exciton (see e.g. Supplementary information in Ref.~\citenum{Trifonov2025arxiv}). This splitting and its axes are defined by the random impurities, composition fluctuations, etc. Averaging of the signal over different splitting energies and/or eigenaxes results in extra factor $\exp{\left( -2\tau_{12}/T_{\delta}^*\right)}$, leading also to significant reduction of the oscillation contrast. Full analytic solution is not practical, instead for moderate magnetic fields $\hbar\Omega_{\pm} \gg \delta_i$ (where $\delta_{1,2}$ are the splittings of the bright exciton) we use phenomenological equations which take into account the effect of fine structure averaging as an extra decay characterized by two times: $T^{\rm QB}_{\rm I}$ and $T^{\rm QB}_{\rm O}$ for ``inner'' and ``outer'' doublets respectively ($T^{\rm QB}_{\rm O}$ also results from fluctuations of dark-bright exciton splitting $\Delta_{\rm X}$). Note that this approach neglects the interplay between these two doublets in the PE signal.

Final equations in the coordinate system related to light propagation ($xyz$) with account on extra dephasing:
\begin{subequations}\label{eq:PEfinSI}
\begin{align}
    \mathbf{P}_{\rm F}^{HH} &= 
        P_0(\tau_{12})\left[
             \mathbf{e}_x \left( 
                1 + \cos(\Omega_{\rm I}\tau_{12})
                \exp\left(-\frac{2\tau_{12}}{T^{\rm QB}_{\rm I}}\right)
            \right)
        +\  \mathbf{e}_y\,\sin(\Omega_{\rm I}\tau_{12})\exp\left(-\frac{2\tau_{12}}{T^{\rm QB}_{\rm I}}\right)
        \right],
    \label{eq-SI:P_HH_F} \\
    \mathbf{P}_{\rm V}^{VV} &= P_0(\tau_{12})\mathbf{e_y} \left[ 1 + \cos(\Omega_{\rm I} \tau_{12})\exp\left(-\frac{2\tau_{12}}{T^{\rm QB}_{\rm I}}\right) \right], 
    \label{eq-SI:P_VV_V} \\ 
\mathbf{P}_{\rm V}^{HH} &= P_0(\tau_{12})\mathbf{e}_x
       \left[
           1 + \frac{\Omega_{\rm X}^{2}}{\Omega_{\rm O}^{2}}            
       + \frac{\Omega_{-}^{2}}{\Omega_{\rm O}^{2}} \cos\!\left(\Omega_{\rm O} \tau_{12} \right)
       \exp\left(-\frac{2\tau_{12}}{T^{\rm QB}_{\rm O}}\right)
       \right]\,.
\label{eq-SI:P_HH_V}
\end{align}\end{subequations}

Note that in experiment, the intensity of interference with original laser signal, $|{\bf P^{\alpha}_{c}}\cdot{\bf E}_l|$, where ${\bf E}_l$ is the electric field of the reference laser pulse. As a result, the absolute value of particular components is measured.

\subsection{Polarization dependence of the PE}
\label{SI-pola}

In Faraday geometry, it is easy to obtain the polarization dependence of the PE signal considering the first pulse linearly polarized along the horizontal ($H$) and the second pulse polarized at an arbitrary angle $\phi$ with respect to it ($R$). Using equations (\ref{eq:rhoGXY_B},\ref{eq:pulse1},\ref{eq:pulse2}), the polarization components of the PE response can be derived:

\begin{equation}
    \mathbf{P}^{HR}_{\rm F} \propto \Gamma_0^2\Bigl[\cos{(2\phi)} + \cos{(\Omega_\text{I}\tau_{12})}\Bigr]\textbf{e}_x
     + \Gamma_0^2\Bigl[\sin{(2\phi)} + \sin{(\Omega_\text{I}\tau_{12})}\Bigr]\textbf{e}_y\,.
    \label{eq:HRH}
\end{equation}
For the orthogonal configurations $\varphi = 0^\circ$ ($HH$ configuration) and $\varphi = 90^\circ$ ($HV$ configuration) the results of previous section are reproduced.

\begin{figure}[H]
    \centering
    \includegraphics[width=0.5\linewidth]{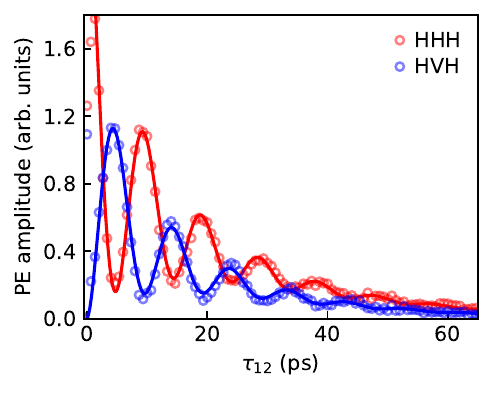}
    \caption{Exemplary PE signals measured in $HHH$ and $HVH$ configurations at $B = \SI{3}{\tesla}$ in Faraday geometry (symbols). The apparent $\pi$-phase shift between the two signals originates from the sign change of the $\cos(\Omega_\mathrm{I}\tau_{12})$ term in Eqs.~\eqref{eq:PFHH} and \eqref{eq:PFHV}. The  lines represent fits based on the respective expressions.}
    \label{fig_s1}
\end{figure}

We consider the situation when the 1st pulse is $H$-polarized. According to Equations~(\ref{eq:PFHH},\ref{eq:PFHV}) when detecting the PE signal in vertical polarization ($V$), both $HHV$ and $HVV$ configurations yield the same component, namely $\frac{1}{2}\sin{(\Omega_\text{I}\tau_{12})}$, as shown for $HHV$ in Figure~\ref{fig_s3}.
Recording the PE signal in horizontal polarization, i.e., in $HHH$ and $HVH$ configurations, a phase shift of $\pi$ between the two cases is expected due to the sign change in the $\cos{(\Omega_\text{I}\tau_{12})}$ term. This phase shift is exemplary shown in Figure~\ref{fig_s1} for a measurement at $B_z = \SI{3}{\tesla}$. Taking the sum of the $HHH$ and $HVH$ signals removes the oscillatory component and yields the pure PE decay $P_0(\tau_{12})$.

\subsection{Data fit}
\label{SI-fit}

\newcommand*{\offset}{A_0 e^{-\sfrac{2\tau_{12}}{\text{T}_0} }}
\newcommand*{\PEdecay}{e^{-\sfrac{2\tau_{12}}{\text{T}_2} }}
\newcommand*{\QB}{\exp{\bigl(\sfrac{-2\tau_{12}}{\text{T}^{ \text{QB} }_{ \text{I} }} \bigr)}}
\newcommand*{\QBO}{\exp{(-\sfrac{2\tau_{12}}{\text{T}^{ \text{QB} }_{ \text{O} }})}}
\newcommand*{\oscprime}{e^{-\sfrac{2\tau_{12}}{\text{T'}_{ \text{QB} }} }}
\newcommand*{\TQBI}{T^\text{QB}_\text{I}}

In the following, we describe the iterative procedure used to determine the $g$-factors of excitons $g_{\rm X}$, electrons $g_\mathrm{e}$, and holes $g_\mathrm{h}$, as well as the excitonic splitting energy between dark and bright states $\Delta_{\rm X}$. Note that the Faraday and Voigt series were not measured at the same sample position and therefore probe slightly different sub-ensembles, which leads to small fluctuation in evaluated values. In each series, the PE-decay at $B=0$~T is measured to determine the coherence times $T_2^\text{s}$ and $T_2$ using Equation~\eqref{eq:expexp} which are set as fixed parameters for fits in magnetic field. The analysis of the PE data follows a stepwise procedure:
\begin{itemize}[nosep]
    \item  First, we determine the oscillation frequency $\Omega$ in the PE decay by applying the corresponding fit Equation~\eqref{eq:PEeqs_F_HHH} for the Faraday geometry, and Equations~\eqref{eq:PEeqs_V_VVV} and \eqref{eq:PEeqs_V_HHH} for the Voigt geometry. 
    \item Second, we extract the $g$-factors from the magnetic field dependence of the oscillation frequency $\Omega_\text{I}(B)$. In Voigt geometry ($HHH$ configuration), the relevant frequency is fitted with $\Omega_\text{O}(B) = \sqrt{\Omega_{\rm X}^2 + \Omega_{-}(B)^2}$. 
    \item For the Voigt data, we independently verify the exchange interaction parameter $\Delta_{\rm X}$ by fitting the magnetic field dependence of the oscillation amplitude  $C=(\Omega_-/\Omega_0)^2$ using the previously determined $g$-factors.
    \item  At the final stage, we reapply the fit equations with fixed $g$-factors and $\Delta_{\rm X}$ fixed, to extract the remaining parameters, including the coherence times $T^\text{QB}$ and the relative  amplitude $R=A^\text{s}/A$.
\end{itemize}
We note that according to our model (see Section~\ref{SI:Theory}) the oscillation amplitude $C$ in Faraday geometry is expected to remain constant with magnetic field strength. However, for $B \lesssim \SI{2}{\tesla}$, $C$ decreases. We attribute such behavior to small splitting of bright exciton states, which limits application of our model in a low magnetic field. In the final step of the iteration, we therefore restrict the application of the model to $B \geq \SI{2}{\tesla}$, where the contrast behaves consistently with the model assumptions. In the preceding steps, however, we include the full data set.

\subsubsection{Faraday geometry}
\label{fit:Faraday}
At $B=0$, the PE signal recorded in the $HHH$ configuration decays with coherence times $T_2^\text{(s)} = \SI{6.3}{\pico\second}$ and $T_2 = \SI{54.3}{\pico\second}$. As mentioned above, the small fluctuations observed, e.g., in comparison with Figure~\ref{fig:0T} of the main text and subsequent data in Voigt geometry, are related to inhomogeneities in the sample. In the $HHV$ configuration, no PE signal is observed at $B = 0$, which is consistent with selection rules for excitonic transitions in this polarization setting. In the $HHV$ configuration, the signal oscillates around zero and changes sign. Because our measurement procedure measures only the amplitude of the PE electric field, an apparent doubling of the oscillation frequency arises that is absent in other polarization configurations. This makes reliable extraction of the fit parameters difficult, particularly at low magnetic fields. We therefore exclude this data set from the fitting procedure. Nevertheless, we show that the parameters extracted from the $HHH$ configuration remain valid and describe the $HHV$ data well.
The fitting procedure is performed using the corresponding expression for the PE signal in the $HHH$ configuration using Equation~\eqref{eq:PEeqs_F_HHH} of the main text.

\begin{figure}[H]
    \centering
    \includegraphics[width=\linewidth]{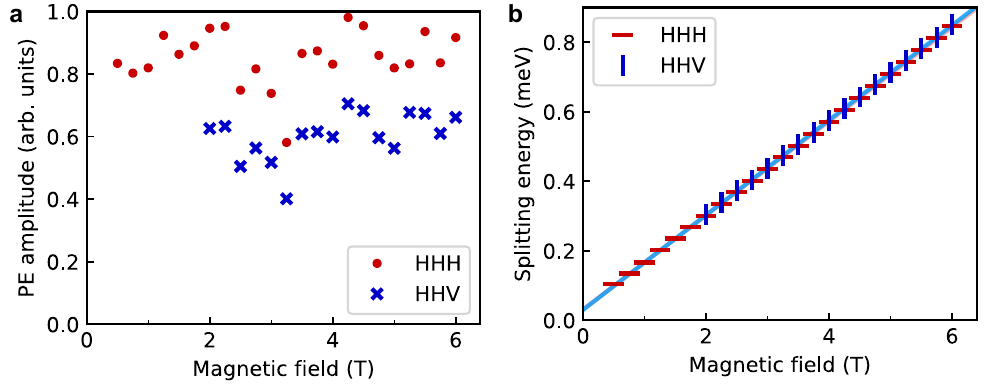}
    \caption{Faraday  geometry. a) The PE amplitudes $A(B)$ for $HHH$ and $HHV$ polarization configurations. For higher magnetic fields ($B \textgreater\SI{2}{\tesla}$), both amplitudes exhibit a similar systematic trend. The scattering is due to adjustment fluctuations. The smaller value in the $HHV$ configuration may be attributed to imperfections in the polarization scheme. b) A linear fit of the splitting energy yields the $g$-factors: $\SI{2.34 \pm 0.03}{} / \SI{2.36 \pm 0.03}{}$ and offsets $\delta$: $\SI{31 \pm 1}{\micro\electronvolt} / \SI{30 \pm 1}{\micro\electronvolt}$, corresponding to the $HHH$/$HHV$ configurations, respectively.}
    \label{fig_s2}
\end{figure}

\paragraph*{1. Determination of the oscillation frequency $\Omega_\text{I}$}\mbox{}\\
In this first iteration step, we focus on extracting the oscillation frequency $\Omega_\text{I}$ from the PE decay over the full magnetic field range from $\SI{0.5}{\tesla}$ to $\SI{6}{\tesla}$. Each transient curve $P_{\rm F}^{HH}(\tau_{12})$ is fitted with the oscillation frequency $\Omega_\text{I}$ and amplitude $A(B)$ treated as free parameters. In addition, the global fit parameters $R$ and $T_\text{I}^\text{QB}$ are determined from data sets with $B \geq \SI{2}{\tesla}$ only. The resulting values $A(B)$ and $\Omega_{\rm I}(B)$ are shown in Figure~\ref{fig_s2} with relative amplitude $R^{HHH}=0.37\pm0.01$ and quantum beats damping $T_\text{I}^{\text{QB},HHH}=42.13\pm0.34$~ps.

The PE amplitudes $A(B)$ in $HHH$ and $HHV$ are expected to be identical. In experiment we observe slightly smaller magnitude in $HHV$ configuration (see Figure~\ref{fig_s2}a). The difference is attributed to limited precision in setting the polarization direction during the measurement. As mentioned above, the PE amplitude in the $HHV$ configuration decreases at small magnetic field strengths. For $B \gtrsim \SI{2}{\tesla}$ the amplitude in the $HHV$ configuration follows the same systematic trend as that observed in the $HHH$ configuration. The overall fluctuations of $A(B)$ are attributed to due to variations in experimental alignment associated with changes in the magnetic field. 

\paragraph*{2. Evaluation of the $g$-factor}\mbox{}\\
To extract the excitonic $g$-factor, we analyze the linear dependence of the splitting energy on the applied Faraday magnetic field strength (see Figure~\ref{fig_s2}b). For this purpose, only data points with $B \geq \SI{2}{\tesla}$ are considered, where the splitting energy exhibits a well-defined linear behavior and the model assumptions are fulfilled. From this fit, we obtain $g = \SI{2.34 \pm 0.03}{} / \SI{2.36 \pm 0.03}{}$ for the $HHH$/$HHV$ configurations, respectively. In addition, a small offset $\delta = 31~\mu$eV is observed. The offset is attributed to additional fine-structure splitting between bright exciton states, fluctuating across localization sites and causing additional dephasing of the oscillations, with $T_\text{I}^{\text{QB},HHH} \approx 2\hbar/\delta = 44$~ps, in good agreement with experiment. This also explains the poor agreement between the experimental data and the existing model at intermediate magnetic fields ($B\lesssim2$~T).

\begin{figure}[H]
    \centering
    \includegraphics[width=0.75\linewidth]{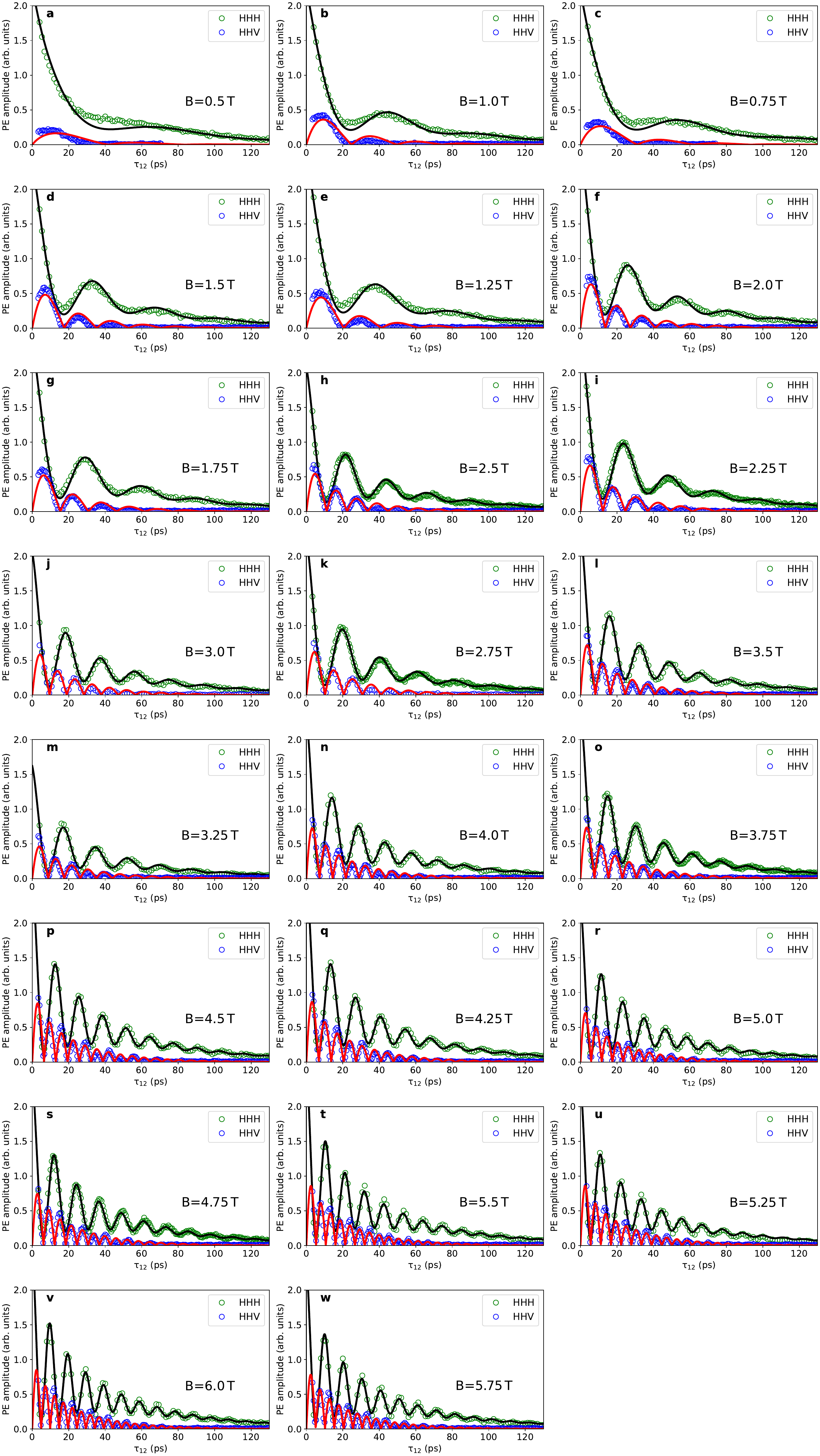}
    \caption{PE measurements in Faraday geometry are shown for magnetic fields ranging from \SIrange[]{0.5}{6.0}{\tesla}, for both polarization configurations: $HHH$ (green) and $HHV$ (blue). The solid lines represent the fit with parameters listed in Table~\ref{table:2faraday}. The data set measured in the $HHV$ configuration is not included in the fitting procedure. Note that the model cannot accurately describe the PE data for $B \lesssim \SI{2}{\tesla}$ due to possible fine structure splitting of bright exciton states with dispersion $\delta \approx 30~\mu$eV.
    }
    \label{fig_s3}
\end{figure}

\paragraph*{3. Applying the model with obtained $g$-factors}\mbox{}\\
In the final iteration step, we apply the fit equations with the previously extracted $g$-factor and include a small correction to account for the offset observed in the linear trend: $\Omega_\text{I} \rightarrow \Omega_\text{I} + \delta$. The model predicts oscillations with constant oscillation amplitude across all magnetic field strengths. However, for small magnetic fields, the experimental data in the $HHH$ configuration exhibit a reduction in oscillation contrast for $B \lesssim \SI{2}{\tesla}$ (see Figure~\ref{fig_s3}). Such behavior is likely due to additional fine structure splitting of bright exciton states, whose optical axes are randomly oriented, giving rise to reduction of oscillation amplitude as discussed in Section~\ref{SI:Theory}. To avoid degrading the extraction of $T_\text{I}^{\text{QB},HHH}$, we therefore restrict the fit to data with $B \geq \SI{2}{\tesla}$. The resulting fit parameters are listed in Table~\ref{table:2faraday} and plotted across the full magnetic field range in Figure~\ref{fig_s3}. The discrepancy between the model prediction and the experimental data at low magnetic fields, particularly visible in the reduced contrast, is also illustrated in this figure. Outside this regime, the signals are in excellent agreement with the model.

\begin{table}[H]
    \centering
    \caption{Overview of free fitting parameter from Equation~\eqref{eq:PEeqs_F_HHH}. Due to variations in experimental alignment associated with changes in the magnetic field, each transient is independently fitted with field-dependent amplitude parameter $A(B)$. Only data set measured at $B \geq\SI{2}{\tesla}$ are considered for parameter A$^\text{(s)}$ and T$_\text{I}^\text{QB}$. }
        \begin{tblr}{colspec={c c c c c}}
        \toprule
        {description} & \SetCell[c=4]{c} fitting parameter \\
        {} & \SetCell[c=2]{c} {$HHH$} && \SetCell[c=2]{c} {$HHV$} \\
        \midrule
        {relative amplitude } &  {$R^{HHH}$}              & {$0.41\pm0.01$}  & {$R^{HHV}$}               & {$=R^{HHH}$}\\
        {quantum beats decay (ps)}              & {$T_\text{I}^{\text{QB},HHH}$}       & {$47\pm0.5$} & {$T_\text{I}^{\text{QB},HHV}$} & {$=T_\text{I}^{\text{QB},HHH}$} \\
        \bottomrule
        \end{tblr}
        \label{table:2faraday}
\end{table}
\newpage

\subsubsection{Voigt geometry}
The PE decay in the $HHH$ configuration at $B=0$ gives coherence times $T_2^\text{(s)} = \SI{8.6}{\pico\second}$ and $T_2 = \SI{56.9}{\pico\second}$. As mentioned above, the small fluctuations observed, e.g., in comparison with Figure~\ref{fig:0T} of the main text and previously discussed data in Faraday geometry, are related to inhomogeneities in the sample. The PE signal recorded in the $VVV$ configuration exhibits oscillations with frequency $\Omega_\text{I}$ and is therefore analyzed analogously to the data measured in the Faraday geometry (see Section~\ref{fit:Faraday}). The decay time of the oscillatory component in the $VVV$ configuration is fixed to $T_\text{I}^\text{QB} = \SI{47}{\pico\second}$, as obtained from the Faraday series. The oscillation frequency in $HHH$ is given by $\Omega_{\text{O}} = \sqrt{\Omega_\text{-}^2 + \BDSS{}^2}$. After determining the difference $g_\text{e} - g_\text{h}$ from the $HHH$ data set, an additional iteration step is performed to verify the excitonic dark-bright splitting parameter $\BDSS{}$. This parameter is extracted exclusively from the amplitude of the oscillation $C=(\Omega_-/ \Omega_\text{O})^2$ in the $HHH$ configuration. The PE transient curves are fitted with Equations~\eqref{eq:PEeqs_V_HHH} and  \eqref{eq:PEeqs_V_VVV}.

\begin{figure}[H]
    \centering
    \includegraphics[width=0.8\linewidth]{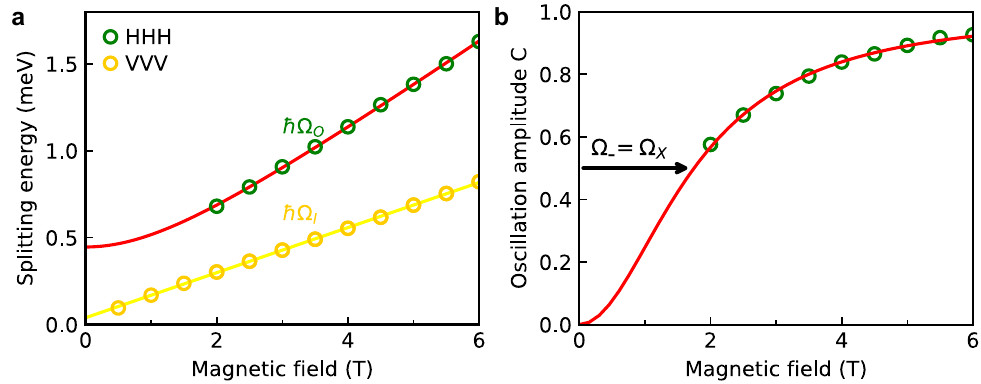}
    \caption{
Voigt geometry. a) Magnetic field dependence of the excitonic splitting energies for the inner $\hbar\Omega_{\rm I}$ and outer $\hbar\Omega_{\rm O}$ doublets extracted from PE signals in $VVV$ (yellow symbols) and $HHH$ (green symbols) configurations, respectively. For $VVV$, the data follow a linear Zeeman-like trend, yielding $g_\text{e} + g_\text{h} = 2.24 \pm 0.04$ and an energy offset $\delta = \SI{38 \pm 4}{\micro\electronvolt}$. The $HHH$ data gives $g_\text{e} - g_\text{h} = 4.52 \pm 0.01$ and $\BDSS{} = \SI{0.45 \pm 0.01}{\milli\electronvolt}$. b) Magnetic field dependence of the amplitude of oscillation $C=(\Omega_{-}/\Omega_{\rm O})^2$ extracted from $HHH$ measurements. The fit uses only the oscillatory prefactor from the $HHH$ model in Equation~\eqref{eq:PEeqs_V_HHH}, with $g_\text{e} - g_\text{h}$ fixed to the value obtained in a), while $\BDSS{}$ remains a free parameter. An arrow marks the point where the magnetic-field-induced splitting matches the intrinsic exchange splitting $\BDSS{}$. This yields $\BDSS{} = \SI{0.46 \pm 0.01}{\milli\electronvolt}$, confirming the result of the frequency-based fit and verifying the robustness of the extracted value.}
    \label{fig_s4}
\end{figure}

\paragraph*{1. Determination of Oscillation Frequencies}\mbox{}\\
The oscillation frequencies $\Omega$ and amplitudes $A(B)$ are treated as free parameters for each individual transient curve. The global fit parameters include the relative amplitude $R^{HHH}=0.80\pm0.01$, $R^{VVV}=0.45\pm0.01$ and the decay constants of the oscillatory contributions, $T_\text{I}^{\text{QB}, VVV}=47$~ps and $T_\text{O}^{\text{QB},HHH}=30$~ps, which are extracted only from data recorded at $B \geq \SI{2}{\tesla}$. The resulting values for the splitting energy $\hbar \Omega_{\rm I}$ and $\hbar \Omega_{\rm O}$ are presented in Figure~\ref{fig_s4}a.

\paragraph*{2. Evaluation of the $g$-factors}\mbox{}\\
The extracted splitting energies $\hbar\Omega$ from the $VVV$ and $HHH$ configurations are plotted against the magnetic field in Figure~\ref{fig_s4}. The $VVV$ data (green symbols) show a clear linear dependence on $B$, consistent with the Zeeman splitting of bright excitons. All data points in this configuration are included in the fit, yielding $g_\text{e} + g_\text{h} = 2.24 \pm 0.02$ and an energy offset $\delta = \SI{38 \pm 4}{\micro\electronvolt}$, in accordance with the data in Faraday geometry and attributed to fine-structure splitting of bright excitonic states. The $HHH$ data (gray symbols) do not allow unambiguous evaluation of the oscillations at $B < \SI{2}{\tesla}$ due to the lower oscillation amplitude, which is also affected by a strong oscillatory signal in the $VVV$ configuration in the presence of slight imperfections of polarization settings. For this reason the data are are present for $B < \SI{2}{\tesla}$. Within this limited field range, the evolution of the splitting energy $\hbar\Omega_\text{O}$ appears nearly linear, and the expected $\sqrt{(\hbar\Omega_-)^2 + \BDSS^2}$ dependence is not clearly resolved. However, the presence of a finite offset in the $HHH$ splitting energy, visible even at high magnetic fields, provides clear evidence for the electron-hole exchange interaction. From the fit, we extract $g_\text{e} - g_\text{h} = 4.52 \pm 0.01$ and $\BDSS = \SI{0.45 \pm 0.01}{\milli\electronvolt}$.

\paragraph*{3. Verification via oscillation amplitude $C(B)$}\mbox{}\\
Since $\BDSS{}$ directly influences the oscillation amplitude in the $HHH$ configuration, we use the magnetic field dependence of $C(B)=(\Omega_{-}/\Omega_{\rm O})^2$ to provide an independent verification of its value. To this end, we re-fit the full data set using Equation~\eqref{eq:PEeqs_V_HHH}, keeping the $g$-factor difference fixed at $g_\text{e} - g_\text{h} = 4.52$, as determined in the previous step. The exchange splitting $\BDSS{}$ and all other parameters remain free. The analysis is again restricted to data with $B \geq \SI{2}{\tesla}$, due to the deviations at lower fields discussed earlier.
Figure~\ref{fig_s4}b shows the extracted oscillation amplitude as a function of magnetic field. From the fit, we obtain $\BDSS{} = \SI{0.46 \pm 0.01}{\milli\electronvolt}$, in very good agreement with the value evaluated from the frequency-based analysis in Figure~\ref{fig_s4}a. This result supports the consistency of the extracted exchange splitting across both fitting approaches.

\begin{figure}
    \centering
    \includegraphics[width=0.5\linewidth]{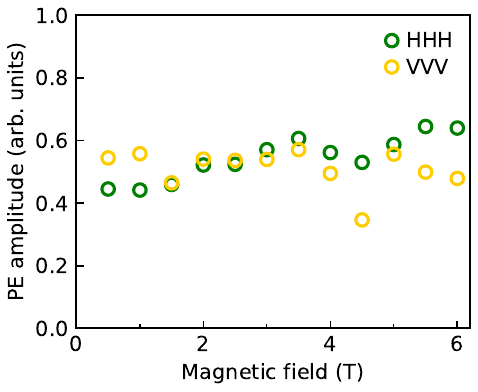}
    \caption{
Magnetic field dependence of the PE amplitudes extracted from fits using fixed $g$-factors and $\BDSS{}$, with the final iteration restricted to data at $B \ge \SI{2}{\tesla}$. The amplitudes correspond to the $VVV$ (green) and $HHH$ (gray) configurations.
}
    \label{fig_s5}
\end{figure}

\paragraph*{4. Final Application of the Model Using Fixed Parameters}\mbox{}\\
\label{fit:final}
In the final iteration step, we apply the model using fixed values for the $g$-factors, $g_\text{e} \pm g_\text{h} = 4.52/2.24$, and the exchange splitting parameter $\BDSS{} = \SI{0.46}{\milli\electronvolt}$, as obtained in the previous steps. As in the case of the Faraday measurements, a correction to the frequency in the $VVV$ configuration is applied via $\Omega_\text{I} \rightarrow \Omega_\text{I} + \delta$. The remaining free fit parameters, amplitudes and coherence times $T_\text{QB}$, are extracted for all data sets with $B \geq \SI{2}{\tesla}$ and are summarized in Table~\ref{table:2voigt} and Figure~\ref{fig_s5}. We further note that the extracted oscillation damping time in the $HHH$ configuration, $T_\text{O}^{{\rm QB,}HHH}$, is relatively short compared to that in the $VVV$ configuration. One plausible cause for this observation is that the exchange splitting parameter \(\BDSS{}\) itself may exhibit a distribution across the probed ensemble of excitons. Such a broadening of the distribution of oscillation frequencies effectively reduces the apparent coherence of the beating signal, leading to a shorter effective damping time. The spread of $\Delta_{\rm X}$ is estimated to be around $2 \hbar /T_\text{O}^{\text{QB},HHH} \approx 60~\mu$eV.

\begin{figure}
    \centering
    \includegraphics[width=0.9\linewidth]{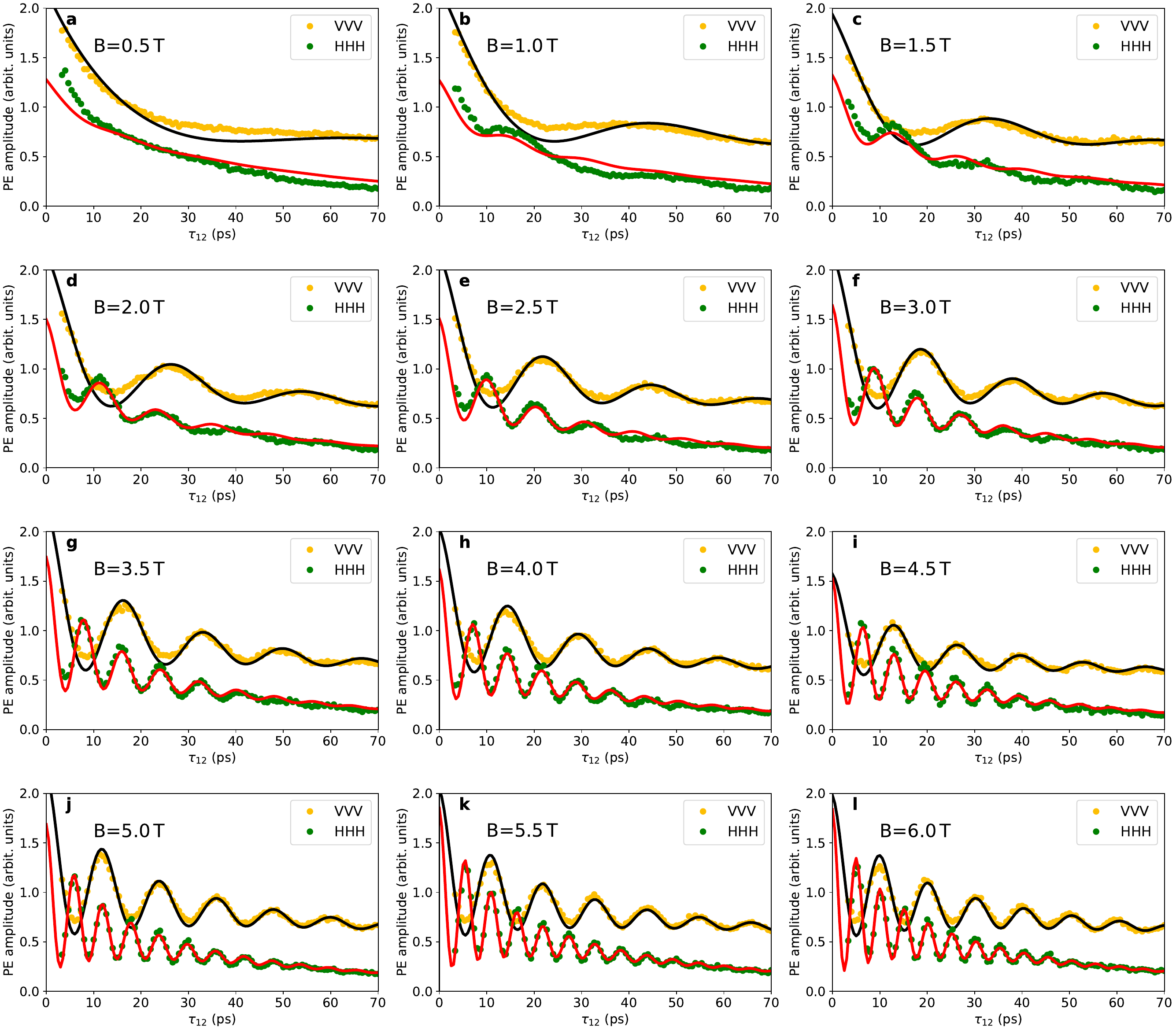}
    \caption{
Measured PE signals (symbols) and corresponding model fits (solid lines) for $VVV$ (yellow) and $HHH$ (green) configurations at selected magnetic field strengths. The fits are based on the final iteration step, using fixed values for the $g$-factors and $\BDSS{}$, as described in Section~\ref{fit:final}. The remaining fit parameters - relative amplitude $R$ and $T_\text{QB}$ — are freely optimized for each data set with $B \geq \SI{2}{\tesla}$. Excellent agreement is observed for both configurations at $B\geq 2$~T. For $HHH$ data below $\SI{2}{\tesla}$, the fit deviates significantly from the measurement, likely due to contributions from additional excitonic transitions not captured by the model.}
    \label{fig_s6}
\end{figure}

Figure~\ref{fig_s6} shows the data and the modeled PE signals across the full magnetic field range. For magnetic fields below $\approx\SI{2}{\tesla}$ in the $HHH$ configuration, the model does not reproduce the data accurately. In this regime, the PE signal exhibits weak oscillations, possibly containing multiple frequency components. A possible explanation is that the signal includes contributions from other polarization configurations resulting from imperfect polarization selection during excitation or detection. We note that our model cannot reliably describe the excitonic states for $B < 2~\text{T}$ due to random splitting of bright excitons with dispersion in the order of $\delta$, which also influences the polarization of the excitonic transitions. Outside this regime, the signals are in excellent agreement with the model.

\begin{table}[H]
    \centering
    \caption{Overview of free fitting parameters in Equations~\eqref{eq:PEeqs_V_HHH} and \eqref{eq:PEeqs_V_VVV}. Due to variations in experimental alignment associated with changes in the magnetic field, each transient is independently fitted with its own amplitude parameter $A(B)$. Only data set measured at $B \geq\SI{2}{\tesla}$ are considered for relative amplitude $R$ and quantum beats damping time $T_\text{O}^\text{QB}$. The value for $T_\text{I}^{\text{QB}, VVV}$ is adopted from measurements in Faraday geometry; see text for details.
    }
        \begin{tblr}{colspec={c c c c c}}
        \toprule
        {description} & \SetCell[c=4]{c} fitting parameter \\
        {} & \SetCell[c=2]{c} {$HHH$} && \SetCell[c=2]{c} {$VVV$} \\
        \midrule
        {amplitude} &  {$A^{HHH}(B)$}              & {fig \ref{fig_s5}} & {$A^{VVV}(B)$} & {fig \ref{fig_s5}}\\
        {relative amplitude} &  {$R^{HHH}$}              & {$0.68\pm0.01$}  & {$R^{VVV}$}   & {$0.55\pm0.02$}\\
        {Quantum beat decay time (ps)}              & {$T_\text{O}^{\text{QB},HHH}$}       & {$21\pm0.4$} & {$T_\text{I}^{\text{QB}, VVV}$} & {47} \\
        \bottomrule
        \end{tblr}
        \label{table:2voigt}
\end{table}

\end{document}